**Modelling multi-cancer screening data to infer on natural history of disease: when can valid, identifiable and precise inference be obtained?**

Modelling multi-cancer screening data

Soares MO[1], Lange J[2], Gogebakan K[3], Dias S[4], Welton NJ[5], Etzioni R[3], Ades AE[5], Palmer S[1]

1 Centre for Health Economics, University of York, York, UK
2 Center for Early Detection Advanced Research, Oregon Health & Science University, Portland, Oregon, US
3 Division of Public Health Sciences, Fred Hutchinson Cancer Center, Seattle, Washington, US
4 Centre for Reviews and Dissemination, University of York, York, UK
5 Population Health Sciences, Bristol Medical School, University of Bristol, UK

**Funding support:** This project is funded by NHS England. The views expressed are those of the authors and not necessarily those of NHS England.
Etzioni was supported by the National Cancer Institute (Award Numbers R35CA274442, U24 CA086368 and UG1 CA286954) and the Rosalie and Harold Rea Brown Chair at the Fred Hutchinson Cancer Center.

**Conflicts:** Lange has undertaken paid consultancy for GRAIL, LLC and Guardant Health, LLC. Other authors have no conflicts to report.

**Abstract**:

Background: Multistate models (MSMs) applied to screening data can characterise the natural history of cancer and predict "stage-shifts" from screening. However, inferring parameters like mean sojourn time (MST) is challenging as disease onset is inherently unobserved in these data. This is even more challenging when characterising heterogeneity between cancer types in multicancer early detection (MCED) trial data.

Methods: We utilised simulated longitudinal MCED screening datasets to evaluate the inferential bounds of MSMs under increasing clinical disaggregation: a 3-state (overall MST), 5-state (early/late stage), and 9-state (stages I-IV) model. Bayesian estimation was performed via Markov chain Monte Carlo. Robustness was assessed through chain convergence, parameter identifiability (via profile likelihood), and precision of estimates. We also explored hierarchical models and the use of informative priors to improve identifiability.

Results: Based only on MCED trial data, many cancer types exhibited inferential challenges. Generally, the 5-state model was as robust as the 3-state model, showing slight improvements to convergence and identifiability while maintaining precision for overall MST. In contrast, the 9-state model showed worsened convergence and identifiability, and a significant reduction in the precision of overall MST estimates. Hierarchical models successfully improved performance, as have informative prior models but the latter introduced bias towards the prior values.

Conclusions: While disaggregating natural history models by individual cancer stages is desirable for policy, these higher-dimensional models show a greater reliance on external data/assumptions. We recommend explicit identifiability assessments and assessments of the influence of external data/assumptions to support inference for MCED screening evaluations.

## 1. Introduction

The modelling of longitudinal cancer screening data using multistate models (MSMs) can characterise the natural history of cancer by describing progression through mutually exclusive states, including 'no cancer' (undetectable cancer), preclinical detectable cancer, and clinical diagnosis.[1] A primary parameter of interest is the (preclinical) mean sojourn time (MST), representing the duration of undetected but detectable cancer in the absence of screening. Where models are disaggregated by cancer stage, the parameters of interest are stage-specific (preclinical) mean sojourn times (stage-MST). These provide a more granular view of progression speeds and help predict a screening test's potential to "stage-shift", that is, to modify the stage distribution at diagnosis.

Independently of the estimation method, whether mathematical[2] or calibration-based[3], inference is technically demanding because the onset of detectable disease is not directly observed. Instead, screening is typically applied in rounds, detecting preclinical disease at a specific time points. Inference is further complicated by the interaction between test sensitivity and sojourn time: a low detection rate may arise from a short MST and high test sensitivity, or a long MST and low test sensitivity. Unique identification requires data on screen-detected and interval cases over multiple screening rounds, and practical identifiability is often limited by data sparsity[4]. To overcome this, analyses often use external evidence (e.g., informative priors) or assumptions (e.g., constraints) to support identification.[5]

These inferential challenges are more significant for multicancer early detection (MCED) tests, such as the Galleri test currently being evaluated in the NHS-Galleri trial (NCT05611632)[6-8]. This trial is powered to detect changes in late-stage cancer diagnosis (TNM stages III and IV), pooled across cancer types. Assessing the population-level impact and cost-effectiveness of MCEDs requires natural history models for distinct cancer types to be informed by the NHS-Galleri trial simultaneously.

This study examines the bounds of inference for natural history modelling of screening data within the MCED context using simulated NHS-Galleri Trial datasets under various structural assumptions. We examine increasing levels of cancer stage disaggregation, starting with a model of overall MST (OMST), then extending it to disaggregate early- and late-stage disease, and finally considering four preclinical and clinical cancer stages. We then examine whether inference can be strengthened and identification improved using external evidence in the form of informative priors or of exchangeability assumptions that allow for flexible information-sharing across cancer types. Our findings can directly inform the specification of the analyses required to support policy/commissioning decisions for the Galleri MCED test, while retaining broader relevance for the evaluation of screening technologies. This work builds on, and advances from, the broader literature that has examined the bounds of inference using practical identifiability assessments in related contexts, including Raue et al (2009) [9] that looked at general partially observed dynamical models, and explorations in applied examples although not in cancer screening, such as Eisenberg and Jain (2018)[10] and Alarid-Escudero et al (2018)[11].

The remainder of this article is structured as follows. Section 2 describes the methods, including the generation of simulated datasets (Section 2.1), the alternative multistate models used to analyse the (simulated) screening data (Section 2.2), their implementation (Section 2.3), and the extensions used to strengthen inference (Section 2.4). Section 2.5 details how we examined generalisability of our findings using a different data simulation model. We then present the results in Section 3 and discuss the implications for future natural history models in Section 4.

## 2. Methods

### *2.1 Generating simulated datasets: Dai's model*

To simulate the NHS-Galleri trial data[6], we used two published models.[12,13] Dai's microsimulation model[12] served as our primary source because it supports the highest level of stage disaggregation (stages I to IV). The alternative model, MCEDsim[13], allowing only disaggregation between early- and late-stage cancer, was used to verify performance under different data generation assumptions (description in Section 2.4).

Dai's model[12] is a microsimulation model based on a cohort of 140,000 individuals of varying ages and gender (ages from 50-77 years old). Then, using US cancer incidence data reduced by 15% (to reflect healthy volunteer effect in clinical trials) and assuming an exponential distribution for the cancer rate, the model simulates, for each individual recruited, the age at clinical detection of cancer, the type of cancer (23 classes) and its stage (stages I-IV). Natural history of preclinical cancer progression was then simulated and applied backwards to clinical detection using stage-specific dwell times. Six scenarios of dwell times were considered.

After generating the simulated study population, individuals are randomly allocated (1:1) to screening or control. The application of three annual screening rounds of screening is simulated using detectable fractions from test sensitivity data from the Circulating Cell-free Genome Atlas (CCGA; NCT02889978) study, reduced by 15% to reflect lower sensitivity in an asymptomatic population.[14] Detectable fractions assume that a cancer detectable at a certain stage will also be detectable at a more advanced stage. Trial mortality was not explicitly considered but a 6% annual dropout rate was applied. For further details, see Dai et al (2024)[12].

Rather than investigating model accuracy, which has been evaluated elsewhere[15,16], we explore the bounds of inference under expected data sparsity. For each dwell time scenario, we selected three datasets reflecting the 2.5th, 50th, and 97.5th percentiles of the (predicted) primary endpoint, totalling 18 datasets. This approach acts as a boundary-condition analysis, evaluating whether the analytical models can achieve robust inference under a range of possible data.

From these patient-level data, we derived aggregated datasets distinguishing counts of screen-detected and interval cases by round—the dataset for the 50th percentile is presented as an example in Table 1 and in Supp A under further disaggregation across cancer stages. For the screening arm, we track counts of individuals screened and of preclinical cancers detected at screens 1, 2 and 3 (n1, 1; n2. r2; n3, r3), alongside counts of interval cancers and person-years of exposure between screens and in the last year of follow-up after the 3rd screen (r4, py4; r5, py5; r6 and py6). Control arm data consists of clinically detected counts and mean exposure time. We collapsed cancer subtypes into 18 different cancer types including a category of 'others'.

<<Table 1 here>>

*2.2 Analysis models: 3-, 5- and 9-state MSM*

Three MSM analysis models considered an increasing level of disaggregation of the preclinical and clinical cancer states into cancer stages: a 3-state model not disaggregating by stage and estimating only OMST, a 5-state model disaggregating early- and late-stage disease, and a 9-state model disaggregating cancer stages I to IV (diagrams in Figure 1). The MSMs define discrete, mutually exclusive states – a brief overview of the models is provided next and a full description in Supp B. Each model includes a "no cancer" state (representing the absence of detectable disease), preclinical screen-detectable states (1, 2 or 4 states, depending on disaggregation), and corresponding clinical disease states. Clinical disease states are absorbing, as the model aims to capture cancer diagnosis rather than post-diagnostic outcomes.

The models are continuous-time Markov models, with exponentially distributed sojourn times. Transition intensities ($qij$) represent transitions rates from state $i$ to state $j$: for example, in the 3-state model, $q23$ is the rate at which a preclinical cancer is clinically identified; its inverse ($1/q23$) reflects the preclinical OMST. See Figure 1 for the parameterisation of the intensity parameters in the different models.

<<Figure 1 here>>

To analyse screening data, we derived transition probability matrices from the intensities using algebraic solutions to the ordinary differential equations, which were generated using Maple software.[17] For instance, across all models, the probability of remaining cancer-free in period of time $t$ is $P11(t) = \exp(-q12t)$. Probabilities for other transitions are more complex and depend on the intensities of both entry and exit from the given state (see Supp B for full description of the probability transition matrices). The models employ a cohort approach, following the initial trial population cohort by applying these transition probabilities and test sensitivities. Screen-detection data are assumed to be binomially distributed, with probability parameters defined by the preclinical cancer prevalence and stage-specific test sensitivity, which is assumed to increase with cancer stage and remain constant across screening rounds. Interval cancers are modelled using a Poisson distribution, where the annual rate is a function of the probability of transitioning to clinical disease and the person-years of exposure. Box 1 presents the transition probability matrix and the 3-state model, while Supp B details the models for the 5- and 9-state models. Derivation of MST is in Figure 1. For 5- and 9- stage models, OMST was predicted from stage-MSTs using progression probabilities, as detailed in Supp A.

<< Box 1 here>>

*2.3 Implementation and criteria for inference*

Analysis models were Bayesian, using weakly informative priors on the intensity parameters to support inference with partially observable data and avoid the informativeness of flat priors on MSTs[18]. For cancer onset intensity ($q_{12}$), we specified a Gamma (0.3, 30) prior, placing 95% of the probability mass between 0.00001 and 0.1, reflecting that clinical incidence in the general population rarely exceeds 0.1. Based on clinical advice that MST would not plausibly exceed 10 years, progression intensities used a Gamma (1.2, 0.9) prior (95% mass: 0.1–5), corresponding to sojourn times between 0.25 and 10 years. Test sensitivities used uninformative Beta (1, 1) priors. The models were estimated in JAGS using R2jags[19], using MCMC, with 100,000 samples (in addition to 50,000 burn-in) across three chains (thinning = 10). The models were initially fitted assuming independent parameters by cancer type (an assumption which will be relaxed in Sections 2.4 and 2.5).

We evaluated robustness of inference based on the following measures and criteria (also summarised in Table 2):

- **Chain convergence** (all parameters): Assessed using the Rhat diagnostic (Gelman Rubin diagnostic). Values < 1.01 indicated excellent convergence, while values above 1.1 indicated no convergence.[20]
- **Identifiability** (MST parameters): refers to whether distinct parameter values correspond to distinct likelihood functions for the observed data[21]. Because estimation is still possible with Bayesian nonidentified models with the posterior being determined entirely or mostly by the prior[22], we assessed whether meaningful information is conveyed in the data, using profile likelihood. We fixed the MST parameter (e.g. $q_{23}$ for the 3-state model) at various values around its point estimate and refitted the model to monitor total residual deviance. To indicate unique identifiability, we looked for a U-shaped plot of total residual deviance according to the (fixed) values of the parameter of interest (the profile likelihood plot), showing lowest model fit statistics when the parameter of interest is close to its point estimate. However, where the remaining parameters adapt to generate similar overall model fits, the profile likelihood plot may instead show a region of significant flattening (indicating practical non-identifiability) or be completely flat (indicating structural non-identifiability).

To determine identifiability we looked for absolute variations in total residual deviance of 3 units, either one sided (partial identifiability) or two-sided (full identifiability). The 3-unit threshold is typically used for the Deviance Information Criteria (DIC)[23].

In 5-state and 9-state models, stage-sojourn parameters were examined using marginal profile likelihood, i.e. the profile likelihood was computed independently for the different parameters determining stage MSTs. To reduce computational burden, profile likelihood was computed by modelling individual cancer type data in all models with independent parameters across cancer types (all except hierarchical models in Section 2.4). Also, the number of MCMC simulations was reduced to (100,000 samples with 100,000 burn-in and thinning of 5). For the complex 9-state model, we used the 10th percentile of the residual deviance distribution to mitigate the against MCMC noise and transient numerical instabilities.

- **Precision** (OMST): evaluated on the MST scale (years) which allows determining whether the width of credible interval covers a sufficiently small range to be clinically useful. Thresholds of 1 and 2 years were used.

We applied the criteria (also in Table 2) to each relevant parameter to classify as fully met, partially met, or not met. As models are multiparametric, the lowest level of each criterion satisfied across relevant parameters by cancer type was summarised; for example, if in the 9-state model identifiability criteria are not met for any one of the four sojourn time parameters the cancer type was classed as not having met identifiability criteria.

<<Table 2 here>>

*2.4 Exploring options to improve identifiability (5-state model)*

For further explorations, we chose the 5-state model because findings are likely generalisable to other levels of disaggregation, and because we are unlikely to obtain robust external evidence (e.g. through expert elicitation) on the four-stage model. To explore potential improvements in identifiability, we adapted the 5-state model using two approaches: a shared-parameter model across cancer types and applying informative prior distributions. Other approaches, data sources and models could have been explored instead; we will return to this in the discussion.

The first approach employed a hierarchical model assuming parameters are exchangeable across cancer types, each drawn from a common population distribution defined by the hyperparameters. To implement this, intensity parameters received Gamma priors and hyperpriors, while test sensitivity parameters received Beta priors and Gamma hyperpriors. To ensure plausibility and structural stability, the priors for intensity parameters were truncated to the same bounds used for the weakly informative priors in previous models (see Box 2 for further detail).

<<Box 2 here>>

The second approach explored the use of informative priors for intensity parameters. We defined hypothetical priors that were intentionally inconsistent with the point estimates from the base model (Section 2.2) to test the trial data's ability to adjust the posterior. Priors for early-stage and late-stage MSTs were centered at 0.5 and 5 years respectively. Precision was defined using effective sample sizes (ESS) of 0.5, 1, 2 and 10 times the (simulated) numbers of cancers detected at early and late stage in the simulated NHS-Galleri trial data. In the discussion, we elaborate on the definition of priors for future analysis, for example, using clinical expert elicitation.

### *2.5 Verification of results with an alternative source of simulated datasets: 5-state model*

To verify the performance of the 5-state analysis model, we used datasets generated from MCEDsim, an R package based on Lange et al (2024)[13] and Gogebakan et al (2025)[24]. MCEDsim uses a natural history of disease model calibrated against UK cancer incidence data from the National Cancer Registration and Analysis Service (NCRAS) for 11 specific cancer types.

The natural history of disease model tracks progression from 'no cancer' to preclinical disease and clinical detection, with cancer states divided into early (I, II) and advanced (III, IV) stages. Overall and late-stage MSTs are specified as Exponential with constant parameter value across cancer types under three alternative scenarios[24]. Cancer-type specific test sensitivities were set at 50% of the values reported in Klein et al (2021)[14]. Time to cancer onset is parameterised through a series of exponential transitions that accommodate an age-increasing hazard function, informed by cancer incidence data and shifted by MST.

MCEDsim simulates cancer natural histories independently across the cancer types modelled, recording the site with the earliest cancer onset and the age and stage at clinical diagnosis. Screening is considered at fixed intervals using site- and stage-specific sensitivities to obtain age and stage at screen diagnosis. For further details of the modelling framework (author communication).

From 200 MCED trial simulated datasets, we selected 9 datasets, corresponding to the 2.5th, 50th and 97.5th percentile of the rate ratio for late-stage cancer in each of the three sojourn time scenarios. As described in Section 2.1, these patient-level simulations were aggregated to inform the 5-state analysis model.

## 3. Results

### *3.1 3-, 5- and 9-state MSM models*

Table 3 presents inference results for MST and stage-MST and the results of explorations of convergence (C), identifiability (I) and precision (P) using a representative dataset (Dai's model, 50th percentile of predicted primary endpoint). Full results, including of test sensitivity, are in Supp C. Across all models, nearly every cancer type show inferential challenges. In the 3-state model, convergence was largely achieved (at least partially), but criteria for identification were not met in 8 (of the 18) cancer types. Precision criteria were not met in 11 (of the 18) cancer types.

The 5-state model predicted OMSTs consistent with those predicted by the 3-state model. The 5-state model, however, showed poorer identification (because two sojourn times need to be identified per cancer type), but slightly better precision for predicted OMST. In most cancers, MST was higher for early-stage than late-stage disease, except for bladder, corpus and uterus, gallbladder, head and neck, liver and melanoma.

Disaggregation to a 9-state model worsened inference across all three criteria. For many cancer types MST estimates showed no clear relationship with stage, and uncertainty over predicted OMST increased. MST estimates in the 9-state model are generally larger than those in the 5-state model, likely because of the wider uncertainty in the modelled parameter (which is the inverse of the OMST). Inference was poorest for prostate cancer (estimate for stage I MST not robust), breast (significant precision lost), and anus, breast and ovary (loss of identifiability).

<< Table 3 here>>

Table 4 summarises performance across all 18 simulated datasets, applying the three criteria cumulatively; we next describe results on whether the three criteria have been satisfied at least partially (C+I+P). Only colorectal, lung and, to a lesser extent, lymphoma satisfied the criteria in most datasets across all models. For breast cancer, criteria were met in most datasets for the 3- and 5-state models, but identification was significantly compromised in the 9-state model. Head and neck, liver and pancreas satisfied this criterion in at least 6 datasets with results being relatively consistent across models. Prostate performed well in the 3-state model, but not in the 5- or 9-state models. Others show a low likelihood of satisfying the criteria, particularly in the 5- and 9-state models. Across datasets, the 3-state model satisfied the three criteria for 5-10 cancer types, compared to 3-9 for the 5-state model and 2-7 for the 9-state model.

<< Table 4 here >>

### *3.2 Exploring options to improve identifiability (5-state model)*

The shared-parameter hierarchical model (Table 5) showed improved performance across all criteria, including significant gains in identifiability and precision. The model with informative priors showed significant improvements in convergence and precision and moderate improvements in identifiability, even at low effective sample sizes in relation to trial data. However, stage-MST estimates show bias towards the prior. Predicted OMSTs proved more robust, likely from additional correlation imposed between parameters. Full results can be found in Supp Tables D2-4.

<< Table 5 here >>

Across datasets (Supp Table D5), the shared-parameter model shows significant improvement in the number of cancer types that satisfy (at least partially) all inference criteria, which range between 12-16 cancer types. The improvements obtained when using imprecise priors are not as marked, ranging between 8 (with lower ESS) and 16 (with medium ESS).

*3.3 Verification of results with an alternative source of simulated datasets*

Applying the 5-state model to MCEDsim datasets (Supp Tables E1-2) confirmed the robustness of these findings. While specific MST estimates varied slightly between simulation sources, OMST results remained similar. The consistency in meeting inferential criteria across both data-generating mechanisms indicates that the limitations of MSMs are robust to different structural assumptions in the underlying simulated data.

## 4. Discussion

Given the inherent heterogeneity in tumor biology, not only in (multi-cancer) test sensitivity but also in preclinical cancer progression, it is essential that analyses of longitudinal cancer screening data reliably estimate mean sojourn time across cancer types and stages to determine the effect of screening on stage shift. In this study, we examined the boundaries of inference when using mathematical multi-cancer multi-state models: specifically, we examined the ability of longitudinal screening data to retrieve valid, identifiable and usefully precise estimates under increasing levels of disaggregation of cancer-stage mean sojourn time, compared to a 3-state model of OMST. Note that, while statistically stable, the 3-state model does not disaggregate by stage and therefore cannot quantify stage-shifts, limiting its utility for broader policy evaluation. Our analyses of simulated multicancer screening datasets showed that:

- a model quantifying MST for late (stages III-IV) and early (stages I-II) stage disease (5-state model) is as robust to data sparsity as the 3-state model, improving convergence and identification and producing OMST predictions with similar precision;
- a model quantifying MST for individual cancer stages from I to IV shows slightly worse convergence and identifiability than the less parameterised 5-state and 3-state models and, importantly, some reduction in the precision of predicted OMST.

Our application revealed that, if the application of the 9-state model is required to allow determining specific stage-shifts (e.g. stage IV to III, III to II etc), external data or assumptions may be required to support inference and identification with longitudinal screening data such as the NHS-Galleri trial.

We explored two approaches to improve identification using the 5-state model (we expect similar conclusions to apply to the 9-state model). The first was a hierarchical shared-parameter model which showed to strengthen inference and improve precision while still allowing the model to reflect heterogeneity between cancer types. The second consisted of using prior distributions on MST, which showed improved convergence and identification but also showed sensitivity to bias, underscoring that while external data can strengthen inference, it needs to be as unbiased as possible. We note that work is currently ongoing that uses a review and structured expert elicitation of UK-based experts to gather beliefs about sojourn times across many cancer types (Jankovic, communication with author). Empirical evidence on test sensitivity could also be used, instead or alongside sojourn time evidence. Such evidence is already available from retrospective studies, like Klein et al (2021)[14]; however, it has been suggested that lower values would be expected in prospective screening setting.[25] Analyses of models for MCED data should seek support for the use of these external data, and for the shared-parameter model assumptions, which may be applied differently to what we did here (for example, across groups of cancer types).

Headline results from the NHS-Galleri trial[26] showed a significant reduction in stage IV diagnoses but not in advanced stage diagnoses (III and IV). This suggests a potential shift within advanced disease (i.e. from stage IV to III). Future work could consider a 6-stage model separating stage III and IV, while grouping stages I and II, which could balance inferential stability.

Our models assume constant test sensitivity across screening rounds; this is the standard in screening modelling. An alternative assumption of fixed 'detectable fractions' across cancer stages, used in Dai's data generating model, still requires stage-specific test accuracy parameters to be defined but further restricts the parameter space by enforcing a monotonic relationship in test sensitivity as cancer progresses. Our results show robustness of convergence, identification and precision to model misspecification (between Dai's and MCEDsim models). However, future work should examine the impact on model fit and model estimates.

Our study formally quantifies the inferential boundaries of multi-state modelling for multi-cancer screening data and is the first to map the limits of applying high-dimensional natural history models across cancer types. We prioritised mathematical-based approaches over calibration because calibration forces convergence to optimised parameter sets without exposing underlying structural uncertainties.[3,27] However, we expect our findings to be applicable to both approaches.

As with the NHS-Galleri trial, future trials of novel multi-cancer early detection (MCED) tests are unlikely to have sufficient sample size to support inferences across every cancer type. Despite variation in design elements (such as number of screening rounds), these trials are therefore likely to yield sparse data for many cancers, including those with lower frequency (e.g. gallbladder), longer latency (e.g. prostate) or low detectability (e.g. melanoma). Consequently, methods to determine the appropriateness of alternative analytical frameworks are essential. Our study provides a methodological framework for multi-state modelling including explicit assessments of (practical) identifiability and examination of the reliability, and influence, of external data and assumptions. Acknowledging this provides a rigorous foundation for future modelling, that reliably evaluates stage-shifts, long-term clinical outcomes, and the cost-effectiveness of emerging screening policies.

Table 1: Exemplar aggregated dataset from Dai's microsimulation model (50th percentile of primary trial endpoint*). Dataset used to inform the 3-state model, not disaggregated by cancer stage.

| | | **Screening arm** | | | | | | | | | | | | **Control arm** | |
|---|---|---|---|---|---|---|---|---|---|---|---|---|---|---|---|
| | | **Screen detected, round 1** | | **Interval detected, between rounds 1 and 2** | | **Screen detected, round 2** | | **Interval detected, between rounds 2 and 3** | | **Screen detected, round 3** | | **Interval detected, after round 3** | | **Detection during follow-up** | |
| | | **r1** | **n1** | **r4** | **py1** | **r2** | **n2** | **r5** | **py2** | **r3** | **n3** | **r6** | **py3** | **SOC** | **py** |
| **All cancers** | | | | | | | | | | | | | | | |
| | | **317** | **70305** | **424** | **63115** | **128** | **60977** | **408** | **58787** | **127** | **56816** | **378** | **54790** | **1988** | **175922** |
| **By cancer type** | | | | | | | | | | | | | | | |
| 1 | Anus | 2 | 70305 | 2 | 63115 | 0 | 60977 | 1 | 58787 | 3 | 56816 | 2 | 54790 | 8 | 175922 |
| 2 | Bladder | 3 | 70303 | 13 | 63114 | 1 | 60975 | 10 | 58786 | 1 | 56812 | 9 | 54788 | 53 | 175909 |
| 3 | Cervix Uteri | 4 | 70300 | 3 | 63107 | 2 | 60961 | 5 | 58781 | 4 | 56801 | 0 | 54781 | 10 | 175855 |
| 4 | Colon and Rectum | 77 | 70296 | 29 | 63106 | 24 | 60956 | 30 | 58778 | 29 | 56792 | 20 | 54781 | 157 | 175840 |
| 5 | Corpus and Uterus | 6 | 70219 | 19 | 63089 | 3 | 60903 | 17 | 58767 | 1 | 56733 | 14 | 54770 | 86 | 175594 |
| 6 | Esophagus | 2 | 70213 | 2 | 63080 | 1 | 60881 | 4 | 58759 | 0 | 56715 | 5 | 54762 | 22 | 175510 |
| 7 | Gallbladder | 1 | 70211 | 2 | 63079 | 2 | 60878 | 2 | 58757 | 2 | 56711 | 3 | 54759 | 9 | 175480 |
| 8 | Head and Neck | 32 | 70210 | 12 | 63078 | 12 | 60874 | 11 | 58756 | 5 | 56707 | 7 | 54758 | 76 | 175466 |
| 9 | Breast cancer | 24 | 70178 | 72 | 63072 | 9 | 60850 | 97 | 58750 | 4 | 56691 | 78 | 54753 | 378 | 175352 |
| 10 | Kidney | 2 | 70154 | 23 | 63032 | 3 | 60769 | 13 | 58705 | 1 | 56590 | 17 | 54708 | 67 | 174937 |
| 11 | Liver | 20 | 70152 | 11 | 63024 | 7 | 60743 | 6 | 58699 | 6 | 56576 | 7 | 54701 | 38 | 174861 |
| 12 | Lung cancer | 73 | 70132 | 56 | 63018 | 34 | 60725 | 41 | 58695 | 45 | 56564 | 38 | 54697 | 315 | 174817 |
| 13 | Lymphoma | 23 | 70059 | 18 | 62988 | 7 | 60635 | 9 | 58676 | 10 | 56478 | 24 | 54676 | 81 | 174414 |
| 14 | Melanoma | 2 | 70036 | 33 | 62977 | 5 | 60610 | 29 | 58670 | 1 | 56459 | 27 | 54662 | 125 | 174309 |
| 15 | Ovary | 7 | 70034 | 6 | 62960 | 6 | 60572 | 10 | 58661 | 1 | 56429 | 6 | 54648 | 28 | 174189 |
| 16 | Pancreas | 20 | 70027 | 7 | 62957 | 3 | 60560 | 18 | 58657 | 7 | 56418 | 10 | 54644 | 54 | 174165 |
| 17 | Prostate | 11 | 70007 | 106 | 62954 | 5 | 60550 | 99 | 58647 | 7 | 56393 | 104 | 54638 | 451 | 174090 |
| 18 | Stomach | 8 | 69996 | 10 | 62899 | 4 | 60439 | 6 | 58601 | 0 | 56287 | 7 | 54590 | 30 | 173647 |

Note that for cancer-site specific counts the numbers screened have been adjusted to reflect conditional independence, i.e. $n_{(i+1)} = n_i - r_i$ .
* the dataset for which the p-value for rate ratio of advanced cancer is the 50th percentile across 100 simulated datasets)

Table 2: Criteria for classifying inference as valid (convergence criteria), identifiable, and precise.

| **Assessment** | **Metric** | **Criteria** | | | **Application** |
|---|---|---|---|---|---|
| | | **Met** | **Partially met** | **Not met** | |
| **Convergence (C)** | Rhat | <1.01 | >1.01, <1.10 | >1.10 | Lowest level reached across all basic model parameters for a cancer type |
| **Identifiability (I)** | Variation in the total residual deviance of models fixing q23 between values of 0.2 and 2. | Threshold value of 3 units crossed twice (two-sided, revealing u-shaped curve): full identifiability | Threshold value of 3 units crossed once (one-sided): partial identifiability | Threshold value of 3 units never crossed. | Lowest level reached across all MST parameters for a cancer type |
| **Precision (P)** | Range of the credible interval | <1 year | >1 year, < 2 years | > 2 years | Level reached on OMST for 3-state model, or on predicted OMST in 5-and 9-state models |

Table 3: 3-, 5- and 9-state models applied to a representative simulated dataset[+]: inference over MST; criteria on chain convergence (C), identifiability (I) and precision (P), classified as met (✓), partially met (|) or not met (✕)

| | | 3-state | | | | 5-state | | | | | | 9-state (mean, 95th CrI) | | | | | | | |
|---|---|---|---|---|---|---|---|---|---|---|---|---|---|---|---|---|---|---|---|
| Cancer type | | OMST | C | I | P | MST e | MST l | OMST | C | I | P | MST I | MST II | MST III | MST IV | OMST | C | I | P |
| 1 | Anus* | 1.1 [0.4,2.9] | ✓ | \| | ✕ | 0.9 [0.3,2.2] | 0.5 [0.1,1.3] | 1 [0.4,2.5] | ✓ | \| | ✕ | 0.8 [0.3,2.2] | 0.3 [0.1,0.8] | 0.3 [0.1,0.8] | 1.1 [0.2,4.4] | 1.2 [0.5,2.7] | ✓ | ✕ | ✕ |
| 2 | Bladder* | 0.7 [0.2,2.3] | \| | ✕ | ✕ | 0.4 [0.1,1.2] | 0.6 [0.2,1.8] | 0.6 [0.2,1.4] | ✓ | ✕ | \| | 0.4 [0.1,1.2] | 0.4 [0.1,1] | 0.3 [0.1,0.9] | 0.3 [0.1,0.8] | 0.8 [0.4,1.7] | ✓ | ✕ | \| |
| 3 | Cervix Uteri | 1 [0.2,3.2] | ✓ | ✕ | ✕ | 0.6 [0.2,1.8] | 0.6 [0.2,1.8] | 0.9 [0.3,2.2] | ✓ | ✕ | \| | 0.5 [0.1,1.4] | 0.5 [0.2,1.6] | 0.4 [0.1,1] | 1.1 [0.2,4.4] | 1 [0.4,2.3] | ✓ | ✕ | \| |
| 4 | Colon and Rectum* | 1.1 [0.7,1.6] | ✓ | ✓ | ✓ | 0.8 [0.5,1.3] | 0.5 [0.3,0.9] | 1.1 [0.7,1.6] | ✓ | \| | ✓ | 0.5 [0.3,0.9] | 0.7 [0.4,1.2] | 0.5 [0.3,0.8] | 0.2 [0.1,0.4] | 1.3 [0.8,1.9] | ✓ | \| | \| |
| 5 | Corpus and Uterus | 0.6 [0.2,2] | \| | ✕ | \| | 0.4 [0.1,1.1] | 0.6 [0.2,1.6] | 0.5 [0.2,1.4] | ✓ | ✕ | \| | 0.4 [0.1,1.1] | 0.4 [0.2,1.2] | 0.4 [0.2,1] | 0.5 [0.2,1.6] | 0.6 [0.3,1.5] | ✓ | ✕ | \| |
| 6 | Esophagus* | 1.6 [0.4,5] | ✓ | ✕ | ✕ | 0.9 [0.3,2.4] | 0.7 [0.2,2] | 1.1 [0.4,2.8] | ✓ | ✕ | ✕ | 0.4 [0.1,1.2] | 0.8 [0.3,1.9] | 0.3 [0.1,0.8] | 0.7 [0.2,2] | 1.2 [0.6,2.5] | ✓ | ✕ | \| |
| 7 | Gallbladder | 1.5 [0.3,5.7] | \| | ✕ | ✕ | 0.6 [0.2,2] | 0.9 [0.2,3.1] | 1 [0.4,2.7] | ✓ | ✕ | ✕ | 0.4 [0.1,1.2] | 0.7 [0.2,1.8] | 0.6 [0.2,1.8] | 0.7 [0.2,2.0] | 1.3 [0.6,2.7] | ✓ | ✕ | ✕ |
| 8 | Head and Neck* | 1.1 [0.6,2] | ✓ | ✓ | \| | 0.5 [0.3,1.1] | 0.8 [0.5,1.4] | 1 [0.6,1.7] | ✓ | \| | \| | 0.6 [0.3,1.3] | 0.3 [0.1,0.5] | 0.6 [0.3,1.1] | 0.3 [0.1,0.5] | 1.2 [0.7,2.1] | ✓ | \| | \| |
| 9 | Breast cancer | 0.4 [0.2,1.1] | ✓ | \| | ✓ | 0.3 [0.1,0.8] | 0.6 [0.2,1.2] | 0.4 [0.2,0.9] | ✓ | \| | ✓ | 0.4 [0.1,1.3] | 0.4 [0.2,0.8] | 0.4 [0.2,0.8] | 0.3 [0.1,0.6] | 0.7 [0.3,1.6] | ✕ | ✕ | \| |
| 10 | Kidney | 0.7 [0.2,2.5] | \| | ✕ | ✕ | 0.4 [0.1,1.2] | 0.5 [0.2,1.4] | 0.5 [0.2,1.4] | ✓ | ✕ | \| | 0.5 [0.1,1.5] | 0.3 [0.1,0.9] | 0.3 [0.1,0.7] | 0.5 [0.2,1.1] | 0.7 [0.3,1.8] | ✕ | ✕ | \| |
| 11 | Liver* | 1 [0.5,2.2] | ✓ | \| | \| | 0.9 [0.4,1.8] | 0.3 [0.1,0.7] | 1 [0.6,2] | ✓ | \| | \| | 0.7 [0.3,1.5] | 0.4 [0.2,0.8] | 0.2 [0.1,0.5] | 0.4 [0.1,0.8] | 1.1 [0.6,2] | ✓ | \| | \| |
| 12 | Lung cancer* | 0.8 [0.6,1.4] | ✓ | ✓ | ✓ | 0.7 [0.4,1.2] | 0.4 [0.2,0.6] | 0.9 [0.6,1.5] | ✓ | \| | ✓ | 0.5 [0.2,0.9] | 0.8 [0.4,1.3] | 0.4 [0.2,0.6] | 0.2 [0.1,0.3] | 1.4 [0.9,2.1] | ✓ | \| | \| |
| 13 | Lymphoma* | 1.2 [0.5,3] | ✓ | \| | ✕ | 0.9 [0.4,2.2] | 0.4 [0.2,0.9] | 1.2 [0.5,2.5] | ✓ | \| | ✕ | 0.7 [0.3,1.7] | 0.6 [0.2,1.2] | 0.4 [0.2,0.7] | 0.2 [0.1,0.5] | 1.4 [0.7,2.6] | ✓ | \| | \| |
| 14 | Melanoma | 0.8 [0.1,3.5] | \| | ✕ | ✕ | 0.4 [0.1,1.2] | 0.6 [0.2,1.8] | 0.5 [0.2,1.3] | ✓ | ✕ | \| | 0.5 [0.1,1.6] | 0.3 [0.1,0.8] | 0.4 [0.1,1] | 0.5 [0.2,1.3] | 0.6 [0.2,1.7] | ✓ | ✕ | \| |
| 15 | Ovary* | 1.6 [0.5,4.5] | ✓ | ✕ | ✕ | 1.2 [0.4,3] | 0.4 [0.2,1.1] | 1.4 [0.5,3.3] | ✓ | \| | ✕ | 0.7 [0.2,1.9] | 0.6 [0.2,1.4] | 0.3 [0.1,0.6] | 0.5 [0.2,1.3] | 1.3 [0.6,2.8] | ✓ | ✕ | ✕ |
| 16 | Pancreas* | 2.2 [0.8,6.1] | ✓ | \| | ✕ | 1.2 [0.5,3.4] | 0.7 [0.3,1.8] | 1.6 [0.7,4.3] | ✓ | ✕ | ✕ | 0.7 [0.2,2.0] | 0.6 [0.3,1.5] | 0.5 [0.2,1.2] | 0.2 [0.1,0.5] | 1.7 [0.8,3.7] | ✓ | ✕ | ✕ |
| 17 | Prostate | 0.5 [0.1,1.6] | ✓ | \| | \| | 0.4 [0.1,1] | 0.4 [0.1,1.1] | 0.4 [0.2,1.1] | ✓ | ✕ | ✓ | 8.3 [0.2,26.8] | 0.7 [0.1,3.4] | 0.3 [0.1,0.7] | 0.5 [0.2,1.1] | 9.0 [0.5,27.4] | \| | ✕ | ✕ |
| 18 | Stomach* | 1 [0.3,2.7] | ✓ | \| | ✕ | 0.6 [0.2,1.6] | 0.6 [0.2,1.6] | 0.8 [0.3,2] | ✓ | ✕ | \| | 0.4 [0.1,1.1] | 0.6 [0.2,1.5] | 0.5 [0.2,1.2] | 0.4 [0.1,0.9] | 1 [0.5,2.1] | ✓ | ✕ | \| |
| **# cancer types for which criteria were fully met, or at least partially met** | | | | | | | | | | | | | | | | | | | |
| ... of all 18 cancer types | | | 13,18 | 3,10 | 3,7 | | | | 18,18 | 0,8 | 4,12 | | | | | | 15,16 | 0,5 | 0,13 |
| ... of 11 (trial) cancer types | | | 10,11 | 3,8 | 2,4 | | | | 11,11 | 0,7 | 2,6 | | | | | | 11,11 | 0,5 | 0,8 |
| **DIC** | | 762 | | | | 1229 | | | | | | 2216 | | | | | | | |

[+] Dataset obtained using Dai's simulation model, dwell time scenario 1, 50%ile of the simulated trial's primary endpoint

* Cancer types selected for the staged analyses as implemented in the NHS-Galleri trial

Table 4: Results of application of the criteria for inference across simulated datasets (Dai's model) and cancer types for the 3-, 5- and 9-state models: number of datasets that achieved criteria* by cancer type, minimum and maximum number of cancer types achieving criteria* across datasets.

| **cancer type** | **3-state model** | | | **5 state model** | | | **9-state model** | | |
|---|---|---|---|---|---|---|---|---|---|
| | **C** | **C+I** | **C+I+P** | **C** | **C+I** | **C+I+P** | **C** | **C+I** | **C+I+P** |
| | Criteria fully met (✓), or at least partially met (✓ or l) out of 18 datasets | | | | | | | | |
| **# datasets where criteria were fully met, or at least partially met for each cancer type** | | | | | | | | | |
| Anus* | 16, 17 | 1, 4 | 0, 0 | 18, 18 | 0, 5 | 0, 0 | 18, 18 | 0, 0 | 0, 0 |
| Bladder* | 10, 17 | 0, 6 | 0, 6 | 18, 18 | 0, 1 | 0, 1 | 18, 18 | 0, 0 | 0, 0 |
| Cervix Uteri | 13, 18 | 0, 7 | 0, 0 | 18, 18 | 0, 6 | 0, 4 | 18, 18 | 0, 0 | 0, 0 |
| Colon and Rectum* | 18, 18 | 18, 18 | 16, 16 | 18, 18 | 1, 18 | 1, 18 | 18, 18 | 0, 18 | 0, 18 |
| Corpus and Uterus | 14, 17 | 0, 7 | 0, 7 | 18, 18 | 0, 3 | 0, 3 | 18, 18 | 0, 1 | 0, 1 |
| Esophagus* | 15, 18 | 2, 6 | 0, 0 | 18, 18 | 1, 2 | 0, 1 | 18, 18 | 0, 1 | 0, 1 |
| Gallbladder | 13, 17 | 0, 1 | 0, 0 | 18, 18 | 0, 0 | 0, 0 | 18, 18 | 0, 0 | 0, 0 |
| Head and Neck* | 18, 18 | 15, 18 | 8, 8 | 18, 18 | 6, 18 | 5, 9 | 17, 18 | 0, 18 | 0, 10 |
| Breast cancer | 18, 18 | 0, 18 | 0, 18 | 18, 18 | 0, 12 | 0, 12 | 13, 17 | 0, 2 | 0, 2 |
| Kidney | 11, 18 | 0, 6 | 0, 6 | 18, 18 | 0, 0 | 0, 0 | 17, 17 | 0, 0 | 0, 0 |
| Liver* | 17, 17 | 10, 16 | 3, 5 | 18, 18 | 0, 16 | 0, 9 | 18, 18 | 0, 12 | 0, 8 |
| Lung cancer* | 18, 18 | 13, 18 | 13, 18 | 18, 18 | 6, 17 | 6, 17 | 18, 18 | 0, 18 | 0, 18 |
| Lymphoma* | 17, 17 | 9, 16 | 7, 11 | 18, 18 | 0, 17 | 0, 13 | 18, 18 | 0, 14 | 0, 12 |
| Melanoma | 7, 17 | 0, 2 | 0, 2 | 18, 18 | 0, 1 | 0, 1 | 16, 18 | 0, 0 | 0, 0 |
| Ovary* | 17, 17 | 8, 15 | 2, 2 | 18, 18 | 0, 13 | 0, 4 | 18, 18 | 0, 3 | 0, 0 |
| Pancreas* | 18, 18 | 7, 18 | 2, 6 | 18, 18 | 1, 13 | 1, 7 | 18, 18 | 0, 10 | 0, 6 |
| Prostate | 16, 18 | 0, 18 | 0, 18 | 18, 18 | 0, 1 | 0, 1 | 5, 17 | 0, 4 | 0, 2 |
| Stomach* | 15, 18 | 1, 11 | 1, 6 | 18, 18 | 0, 4 | 0, 4 | 18, 18 | 0, 1 | 0, 1 |
| **min - max# cancer types for which criteria were fully met, or at least partially met, across datasets** | | | | | | | | | |
| | **11-17, 16-18** | **2-7, 8-16** | **1-5, 5-10** | **18-18, 18-18** | **0-4, 4-11** | **0-4, 3-9** | **15-18, 16-18** | **0-0, 3-8** | **0-0, 2-7** |

* Criteria considered were: chain convergence (C), chain convergence and identifiability (C+I) and chain convergence, identifiability and precision (C+I+P).

Table 5: Shared-parameter 5-state model applied to a representative simulated dataset[+]: criteria on chain convergence (C), identifiability (I) and precision (P) classified as met (✓), partially met (|) or not met (✕)

| | | MST e | MST l | OMST | C | I | P |
|---|---|---|---|---|---|---|---|
| **1** | Anus* | 0.6 [0.3,1.1] | 0.3 [0.2,0.6] | 0.7 [0.3,1.3] | ✓ | \| | ✓ |
| **2** | Bladder* | 0.3 [0.2,0.6] | 0.3 [0.2,0.7] | 0.4 [0.2,0.7] | ✓ | \| | ✓ |
| **3** | Cervix Uteri | 0.6 [0.2,1.1] | 0.3 [0.2,0.7] | 0.7 [0.3,1.4] | ✓ | ✕ | \| |
| **4** | Colon and Rectum* | 0.7 [0.5,1.1] | 0.4 [0.2,0.7] | 0.9 [0.6,1.4] | ✓ | \| | ✓ |
| **5** | Corpus and Uterus | 0.3 [0.2,0.5] | 0.4 [0.2,0.8] | 0.4 [0.2,0.7] | ✓ | \| | ✓ |
| **6** | Esophagus* | 0.5 [0.2,1.1] | 0.4 [0.2,0.8] | 0.6 [0.3,1.2] | ✓ | \| | ✓ |
| **7** | Gallbladder | 0.5 [0.2,1.0] | 0.4 [0.2,0.8] | 0.6 [0.3,1.2] | ✓ | ✕ | ✓ |
| **8** | Head and Neck* | 0.6 [0.3,1.1] | 0.7 [0.4,1.3] | 1.0 [0.6,1.7] | ✓ | \| | \| |
| **9** | Breast cancer | 0.2 [0.2,0.4] | 0.4 [0.2,0.7] | 0.3 [0.2,0.4] | ✓ | \| | ✓ |
| **10** | Kidney | 0.3 [0.2,0.6] | 0.3 [0.2,0.6] | 0.4 [0.2,0.7] | ✓ | \| | ✓ |
| **11** | Liver* | 0.7 [0.4,1.3] | 0.3 [0.2,0.6] | 0.8 [0.5,1.4] | \| | \| | ✓ |
| **12** | Lung cancer* | 0.8 [0.4,1.3] | 0.4 [0.2,0.6] | 1.0 [0.6,1.6] | ✓ | \| | ✓ |
| **13** | Lymphoma* | 0.8 [0.4,1.5] | 0.3 [0.2,0.6] | 1.0 [0.5,1.7] | ✓ | \| | \| |
| **14** | Melanoma | 0.2 [0.2,0.3] | 0.4 [0.2,0.7] | 0.3 [0.2,0.4] | ✓ | \| | ✓ |
| **15** | Ovary* | 0.8 [0.4,1.6] | 0.3 [0.2,0.6] | 1.0 [0.5,1.8] | ✓ | \| | \| |
| **16** | Pancreas* | 0.9 [0.5,1.6] | 0.5 [0.2,0.9] | 1.1 [0.7,2.1] | ✓ | \| | \| |
| **17** | Prostate | 0.2 [0.2,0.4] | 0.3 [0.2,0.6] | 0.3 [0.2,0.4] | \| | \| | ✓ |
| **18** | Stomach* | 0.5 [0.3,1.0] | 0.4 [0.2,0.8] | 0.7 [0.3,1.2] | ✓ | ✕ | ✓ |
| **# cancer types for which criteria were fully met, or at least partially met** | | | | | | | |
| # ✓, # ✓or \|, out of all 18 cancer types | | | | | 16,18 | 0,15 | 13,18 |
| # ✓, # ✓or \|, out of the 11 (trial) cancer types with * | | | | | 10,11 | 0,10 | 7,11 |
| **DIC** | | 1258 | | | | | |

[+] Dataset obtained using Dai's simulation model, dwell time scenario 1, 50%ile of the simulated trial's primary endpoint
* Cancer types selected for the staged analyses as implemented in the NHS-Galleri trial

Figure 1: Diagrams for 3-state, 5-state and 9-state models.

**3-state model**

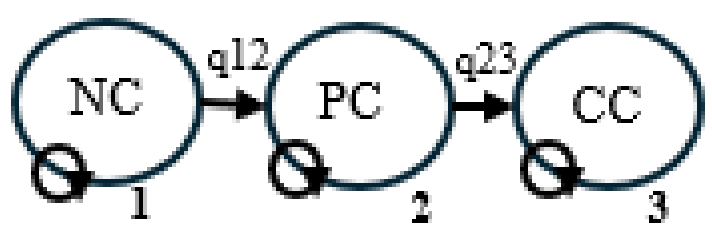

NC: no cancer
PC: pre-clinical cancer
CC: clinical cancer
qij: transition intensity between states i and j

Natural history parameters: q12,q23

Quantities of interest for inference*:
Preclinical MST = 1/q23

**5-state model**

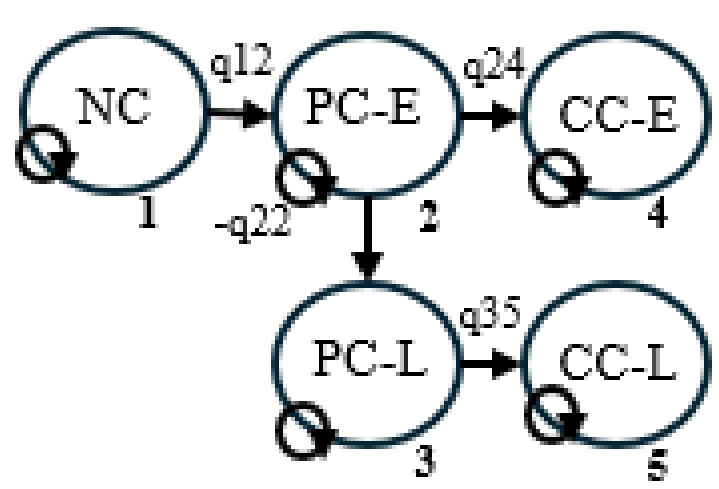

NC: no cancer
PC-E: pre-clinical early-stage cancer
CC-E: clinical early-stage cancer
PC-L: pre-clinical late-stage cancer
CC-E: clinical late-stage cancer
qij: transition intensity between states i and j
Natural history parameters: q12,q24,q22,q35

Quantities of interest for inference*:
Preclinical EMST = 1/q22
Preclinical LMST = 1/q35

**9-state model**

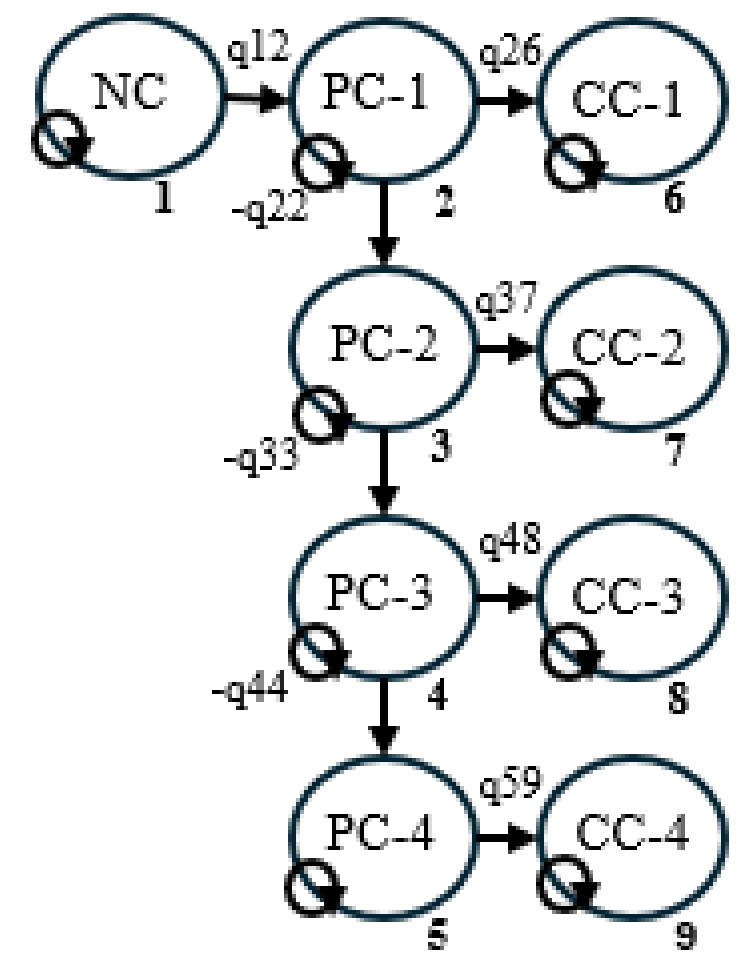

NC: no cancer
PC-i: pre-clinical stage i of cancer
CC-i: clinical stage i of cancer
qij: transition intensity between states i and j
Natural history parameters: q12,q26,q22,q37,q33,q48,q44,q59

Quantities of interest for inference*:
Preclinical MST in stage 1: 1/q22
Preclinical MST in stage 2: 1/q33
Preclinical MST in stage 3: 1/q44
Preclinical MST in stage 4: 1/q59

* the quantities of interest for inference are derived assuming exponential distributions for NHD parameters

Box 1: Specification of the 3-state model

**Intensity matrix**

| | 1 | 2 | 3 |
|---|---|---|---|
| 1 | $-q12$ | $q12$ | 0 |
| 2 | 0 | $-q23$ | $q23$ |
| 3 | 0 | 0 | 0 |

**Probability transition matrix for time interval t**

| | 1 | 2 | 3 |
|---|---|---|---|
| 1 | $P11(t)$ | $P12(t)$ | $P13(t)$ |
| 2 | 0 | $P22(t)$ | $P23(t)$ |
| 3 | 0 | 0 | 1 |

Elements of the probability transition matrix defined as a function of elements of the intensity matrix

$P11(t) = e^{-q12 \cdot t}$

$P12(t) = \frac{q12}{q12 - q23} \cdot (e^{-q23 \cdot t} - e^{-q12 \cdot t})$

$P13(t) = 1 - e^{-q12 \cdot t} - \frac{q12}{q12 - q23} \cdot (e^{-q23 \cdot t} - e^{-q12 \cdot t})$

$P22(t) = e^{-q23 \cdot t}$

$P21(t) = 1 - P22(t)$

**Model based on a cohort approach**

The model follows a cohort of individuals recruited into the trial – i.e. at t=0 (at time 0) and s=0 (before screening).

In this initial cohort, the presence of (undetected) pre-clinical cancer is unobserved and is defined as a function of the transition probabilities under an assumed duration of cancer in those with cancer at t=0 given by a, as follows:

$Pr[S = 2|\ t = 0, s = 0] = P12(a)/(P12(a) + P11(a))$

The proportion without cancer is :

$Pr[S = 1|\ t = 0, s = 0] = 1 - Pr[S = 2|\ t = 0, s = 0]$

Screen detection in the first round is described as

$r_1 \sim Bin(p_1, n_1)$

with probability parameter described as a function of test sensitivity, *sens*, applied to the proportion with pre-clinical cancer:

$p_1 = sens \cdot Pr[S = 2|t = 0, s = 0]$

Detected cases are not followed up further for detection, and therefore the proportion left with pre-clinical cancer after screening (s=1) includes only those undetected, as follows

$Pr[S = 2|\ t = 0, s = 1] = (1 - sens) \cdot Pr[S = 2|t = 0, s = 0]$.

The proportion with no cancer being followed up remains unchanged

$Pr[S = 1|\ t = 1, s = 0] = Pr[S = 1|\ t = 0, s = 0]$.

Interval cancers between the first and second screens are modelled as

$r_4 \sim Poisson(\lambda_1)$

$\lambda_1 = p_4 \cdot py_4/t$

$p_4 = Pr[S = 1|\ t = 0, s = 1] \cdot P13(t) + Pr[S = 2|\ t = 0, s = 1] \cdot P23(t)$

reflecting the probability that those without cancer develop clinically detected cancer over that period, and the probability that those with pre-clinical cancer show clinical detection within the same period.

Before the second screen at t=1 and s=0, the interval cases are removed, and the cohort distribution is as follows:

$Pr[S = 1|\ t = 1, s = 0] = Pr[S = 1|\ t = 0, s = 1] \cdot P11(t)$

$Pr[S = 2|\ t = 1, s = 0] = Pr[S = 1|\ t = 0, s = 1] \cdot P12(t) + Pr[S = 2|t = 0, s = 1] \cdot P22(t)$

Screen detection at the 2nd screening round is described as:

$r_2 \sim Bin(p_2, n_2)$

$p_2 = sens \cdot Pr[S = 2|t = 1, s = 0]$

After screening (s=1), the cohort is updated as follows:

$Pr[S = 1|\ t = 1, s = 1] = Pr[S = 1|\ t = 1, s = 0]$

$Pr[S = 2|\ t = 1, s = 1] = (1 - sens) \cdot Pr[S = 2|t = 1, s = 0]$

Interval cancers between the first and second screens are modelled as:

$r_5 \sim Poisson(\lambda_2)$

$\lambda_2 = p_5 \cdot py_5/t$

$p_5 = Pr[S = 1|\ t = 1, s = 1] \cdot P13(t) + Pr[S = 2|\ t = 1, s = 1] \cdot P23(t)$

Before the third screen at t=2 and s=0, the interval cases are removed, and the cohort distribution as follows:

$Pr[S = 1|\ t = 2, s = 0] = Pr[S = 1|\ t = 1, s = 1] \cdot P11(t)$
$Pr[S = 2|\ t = 2, s = 0] = Pr[S = 1|\ t = 1, s = 1] \cdot P12(t)\ +\ Pr[S = 2|t = 1, s = 1] \cdot P22(t)$

Screen detection at the 3rd screening round is described as:
$r_3 \sim Bin(p_3, n_3)$
$p_3 = sens \cdot Pr[S = 2|t = 2, s = 0]$

After screening (s=2), the cohort is updated as follows
$Pr[S = 1|\ t = 2, s = 1] = Pr[S = 1|\ t = 2, s = 0]$
$Pr[S = 2|\ t = 2, s = 1] = (1 - sens) \cdot Pr[S = 2|t = 2, s = 0]$

Interval cancers 12 months after the 3rd screen are described as:
$r_6 \sim Poisson(\lambda_3)$
$\lambda_3 = p_6 \cdot py_6/t$
$p_6 = Pr[S = 1|\ t = 2, s = 1] \cdot P13(t) +\ Pr[S = 2|\ t = 2, s = 1] \cdot P23(t)$

Detection in the control group consists only of clinical detection (no screening implemented) and is therefore described as:
$r_c \sim Poisson(\lambda_c)$
$\lambda_c = p_c \cdot py_c/t$
$p_c = Pr[S = 1|\ t = 0, s = 0] \cdot P13(t.c) +\ Pr[S = 2|\ t = 0, s = 0] \cdot P23(t.c)$

**Priors**
$q_{12} \sim Gamma(0.3{,}30)$
$q_{23} \sim Gamma(1.2{,}0.9)$
$sens \sim Beta(1{,}1)$

Box 2: Specification of the random effects across cancer types for the sharing of information model

**Priors by cancer type, *i***

$q_{12,i} \sim Gamma(A1, B1)$, $q_{12,i} \in [0.0001, 0.1]$

$delta_i \sim Gamma(A2, B2)$, $delta_i \in [0.1, 5]$

$q_{22,i} = q_{24,i} + delta_i$

$q_{23,i} \sim Gamma(A3, B3)$, $q_{23,i} \in [0.1, 5]$

$q_{35,i} \sim Gamma(A4, B4)$ , $q_{35,i} \in [0.1, 5]$

$sens.e_i \sim Beta(Asens.e, Bsens.e)$, $sens.e_i \in [0.1, 0.99]$

$ratio_i \sim Beta(Aratio, Bratio)$ , $ratio_i \in [0.1, 0.99]$

$sens.l_i = sens.e_i + (1 - sens.e_i) * ratio_i$

**Hyperpriors**

$A1 \sim Gamma(1, 0.1)$

$B1 \sim Gamma(0.1, 0.01)$

$A2 \sim Gamma(1, 0.1)$

$B2 \sim Gamma(0.1, 0.01)$

$A3 \sim Gamma(1, 0.1)$

$B3 \sim Gamma(0.1, 0.01)$

$A4 \sim Gamma(1, 0.1)$

$B4 \sim Gamma(0.1, 0.01)$

$Asens.e \sim Gamma(1, 0.1)$

$Bsens.e \sim Gamma(0.1, 0.01)$

$Aratio \sim Gamma(1, 0.1)$

$Bratio \sim Gamma(0.1, 0.01)$